# *Plasmonic sensing using Babinet's principle*

*Joseph Arnold Riley*[1,2,#], *Michal Horák*[3,#], *Vlastimil Křápek*[3,*], *Noel Healy*[1], *and Victor Pacheco-Peña*[1,*]

[1]*School of Mathematics, Statistics and Physics, Newcastle University, Newcastle Upon Tyne, NE1 7RU, United Kingdom*
[2]*School of Engineering, Newcastle University, Newcastle Upon Tyne, NE1 7RU, United Kingdom*
[3]*Central European Institute of Technology, Brno University of Technology, Purkyňova 123, 612 00, Brno, Czech Republic*
[4]*Institute of Physical Engineering, Brno University of Technology, Technická 2, 616 69, Brno, Czech Republic*

**Developing methods to sense local variations in nearby materials, such as their refractive index and thickness, is important in different fields including chemistry and biomedical applications, among others. Localized surface plasmons (LSPs) excited in plasmonic nanostructures have demonstrated to be useful in this context due to the spectral location of their associated resonances being sensitive to changes near the plasmonic structures. In this manuscript, Babinet's principle is explored by exploiting LSP resonances excited in complementary metal-dielectric cylindrical plasmonic structures (plasmonic particle-dimers and aperture-dimers in our case). Both plasmonic structures are evaluated numerically and experimentally using Electron Energy Loss Spectroscopy (EELS), providing a full physical understanding of the complementary nature of the excited LSP resonances. The studied plasmonic structures are then exploited for dielectric sensing under two configurations: when a thin dielectric film is positioned atop the plasmonic structures and when the analyte surrounds/fills the plasmonic particles/apertures. The complementary sensing performance of both proposed structures is also evaluated, showing the approximate validity of the Babinet principle with sensitivities values of up to 700 nm/RIU for thin dielectric sensing.**

*# These authors contributed equally to this work*
[*]*email: victor.pacheco-pena@newcastle.ac.uk, vlastimil.krapek@ceitec.vutbr.cz*

# Introduction

Metallic nano-/micro-structures acting as plasmonic antennas have been the focus of intense research in recent decades [1]–[4]. This is due to their ability to confine and strongly enhance incident electromagnetic (EM) fields at sub-wavelength scales in the visible and near-infrared spectral range [5]–[7]. The spatial confinement of EM radiation is a result of coupling between the EM field and conduction electrons in metallic structures, provided the dispersive nature of metals at optical frequencies [1], [8], [9]. The manipulation and control of light-matter interactions at the nano-/micro-scale has been at the core of the field of plasmonics with emphasis in surface plasmon polaritons (SPPs, evanescent surface waves that are excited at the interface between a metal and a dielectric) [10]–[17] and localized surface plasmon (LSP) resonances excited in plasmonic nanoparticles (such as bowtie nanoantennas, nanospheres, disks, nanocrescents, among others) [18], [19], [28], [29], [20]–[27]. The focus of this manuscript is on LSP resonances appearing in plasmonic nanostructures.

The spectral location of the LSP resonances of a plasmonic structure can be controlled by engineering the geometrical parameters of the involved nanoparticles such as size and shape (spheres, nanorods, triangular monomers [20], [30]–[36], among others) as well as the materials used to design them, including the metal and dielectrics used [33], [34], [37]–[39]. Importantly, the excitation of LSP resonances in plasmonic structures can give rise to localized hotspots [21], [30], [40]–[44]. This ability to confine EM radiation at the nanoscale and the arbitrary control of the spectral location of a LSP resonance has enabled plasmonic nanostructures to be implemented in new and improved applications such as photo-catalysis [45], optical trapping [41] and data storage [46]. The field of sensing has also benefited from the research and innovation of plasmonic nanostructures with some examples including single particle detection [47]–[49], surface-enhanced Raman spectroscopy [50], [51], biosensors [52]–[55] as well as the ability to determine the optical/geometrical properties of local dielectrics (such as the refractive index or their thickness) [56], [57]. In the latter example, material characterization can be accomplished by measuring the change of the spectral location of LSP resonances excited in plasmonic nanostructures when a dielectric analyte with refractive index $n_a$ and thickness $\delta_a$ is placed in their vicinity.

As it is known, LSP resonances in plasmonic nanostructures can also be engineered by combining the resonances that appear in different nanoparticles. For instance, exploiting single nanoantennas to design dimers and trimers or arrays of particles [18], [30], [31], [40], [41], [44], [58]–[60]. Classical concepts have also been implemented in plasmonics to understand and design plasmonic nanostructures, such as the study of LSP resonances using conformal mapping (a technique known in antenna engineering [24], [61]–[63]). In this context, Babinet's principle for complementary structures is another example of classical techniques that has recently been successfully studied in plasmonic structures [40], [64]–[69]. Babinet's principle states that the diffraction pattern of an infinitely thin set of perfectly opaque particles will be identical to a complementary set of apertures in an infinitely thin, perfectly opaque screen with only a difference in the observed amplitude [64]. Recently, several studies have demonstrated the possibility of exploiting this method in the field of plasmonics where plasmonic nanoparticles will have their respective LSP resonances almost at the same spectral location as their complementary versions [30], [40], [59], [65]–[68]. It is important to note that small discrepancies will be expected [67] in terms of the spectral location of LSP resonances due to, for instance, metals not being perfect conductors at optical frequencies and also the fact that the thickness of plasmonic nanostructures, although small, is not almost zero as required by the classical Babinet's principle [70].

In this manuscript, we conduct an in-depth numerical and experimental study on the performance of plasmonic cylindrical particle-dimers made of gold (from now on just referred to as plasmonic particles) and their complementary version, cylindrical aperture-dimers in a metallic screen (from now referred to as plasmonic apertures), to investigate the validity of Babinet's principle for such structures. Electron Energy Loss Spectroscopy (EELS) is used to experimentally map the distribution of the LSP modes, and these results are compared with numerical simulations using the transmission/reflection spectra as well as a full study of the field distribution (electric and magnetic fields) of their corresponding LSP resonances. Additionally, as both plasmonic structures confine EM field at nanoscales, we carry out a study on their performance when they are used as dielectric sensing devices. To do this, two different configurations are studied for the analyte to be sensed: i) when a dielectric analyte is positioned atop the plasmonic structures and ii) when the dielectric analyte is surrounding the plasmonic particles or filling the plasmonic apertures. It will be shown how the complementary plasmonic apertures performed favorably in terms of sensitivity over the plasmonic particles when the analyte is positioned atop the structures while the opposite is true when the analyte is used in the second configuration, achieving sensitivity values in the order of hundreds of nm/RIU.

**Design and Results**

**Configurations of the plasmonic particles and apertures**

Three-dimensional (3D) schematic representations of the two plasmonic structures under study are shown in Fig. 1a,d: particles and apertures, respectively. Two-dimensional (2D) cross-sections on the *yz*-plane of these plasmonic structures are also shown in Fig. 1b for the plasmonic particles and Fig. 1e for the plasmonic apertures, for completeness. As observed, the plasmonic structures are complementary to one another. The plasmonic particles are made of gold (Au) with a thickness of 30 nm and the complementary version (plasmonic apertures) are made into an Au sheet of the same thickness. In so doing, the layer thickness is smaller than the incident wavelength of the illuminating planewave which will be varied between $\lambda_0$ = 0.967 µm to ~2 µm (310 – 150 THz), with $\lambda_0$ as the wavelength in free space. This is to enable an approximation of Babinet's principle (which considers infinitely thin metallic particles/screens). The plasmonic particles and apertures are characterized by a diameter *D* and separation between the particles/apertures defined by *L*. Both structures are placed on top of a silicon nitride ($Si_3N_4$) substrate. To study the plasmonic structures, in the numerical analysis, Au is modelled by fitting Johnson and Christy's experimental data [71]. The substrate, $Si_3N_4$, is modelled using the experimental work from [72]. The structures are illuminated with a planewave under normal incidence traveling along the *z*-axis. As it will be shown later, the linear polarization of the incident illumination will be changed to be along the *x*- or *y*-axis to ensure the validity of Babinet's principle (as orthogonal polarizations are required to excite complementary modes in particles and apertures [30], [67]).

As it will be shown below, we will consider thin dielectric analytes on top of the plasmonic structures to evaluate their performance as dielectric sensors. The schematic representation of this setup is also shown in Fig. 1. This configuration will be modified and discussed in later sections to consider other sensing scenarios such as the case when the analyte surrounds the plasmonic particles or fills the plasmonic apertures. Without loss of generality, to evaluate the sensing performance of the designed structures, we consider dielectric analytes as being non-dispersive within the spectral range under study having real refractive index values ($n_a$) and variable thickness ($\delta_a$) along the *z*-axis (see Fig. 1). Finally, the whole structures are immersed in air ($n_0$ = 1).

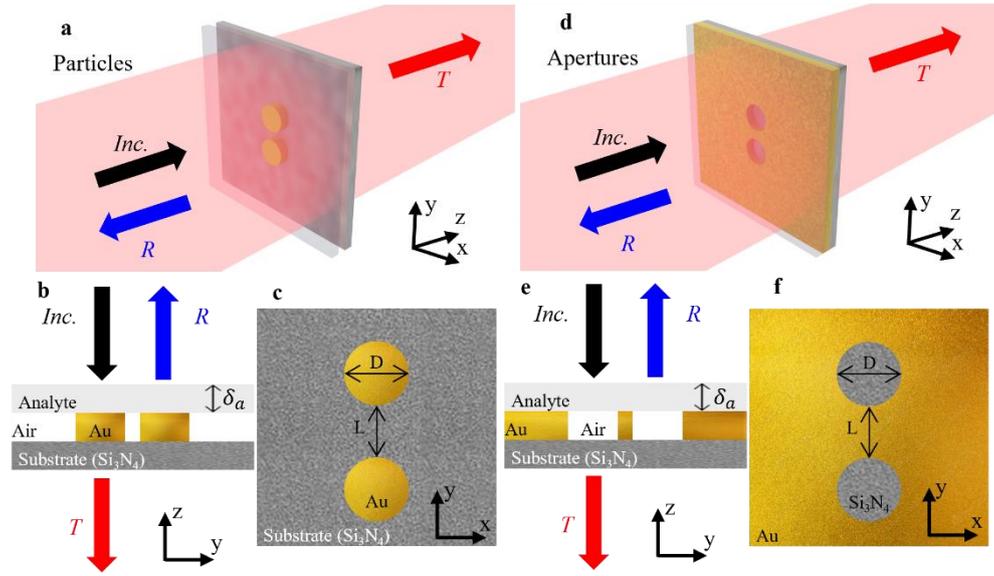

**Fig. 1| Schematic representation of the complementary plasmonic structures.** (a-c) Cylindrical gold (Au) particle-dimers on a silicon nitride ($Si_3N_4$) substrate, (a) perspective, (b) cross section on the *yz*-plane and (c) *xy*-plane (d-f) Same as (a-c) but for the complementary plasmonic structure consisting of cylindrical aperture-dimers in an Au film. A dielectric with variable thickness and refractive index is positioned atop these metallic structures which will act as the analyte to be sensed. The two structures are illuminated by an incident planewave, represented by the black arrow (*Inc.*), this illumination enables the excitation of LSP resonances. The spectral position of these LSP resonances for each plasmonic structure is determined using the reflection (*R*) and transmission (*T*) coefficients. These parameters are then used to study the sensing capabilities of both plasmonic structures in the presence of the analyte. *D* and *L* in (c,f) represent the diameter and the separation of the cylindrical components, respectively.

For the sake of completeness, the *xy*-plane at the top surface of the plasmonic particles and apertures are shown in Fig. 1c,f, respectively. As mentioned in the previous section, the size and shape of the plasmonic structures will dictate the spectral location of the LSP resonances. Different studies have been recently reported in relation to the effect of changing geometrical parameters. For instance, modifying *D* and/or *L* results in a shift in the LSP resonance wavelength [42], [73], as expected. Based on this, for the numerical studies we will consider these parameters to be constant (*D* = 200 nm, *L* = 20 nm) in this manuscript and focus our attention into studying the validity of Babinet's principle with the proposed designs and their performance as dielectric sensors.

**Numerical setup**

The proposed plasmonic structures shown in Fig. 1 are numerically studied using the RF solver of the commercial software COMSOL Multiphysics®. The full plasmonic structures (particles or apertures on top of a $Si_3N_4$ substrate of thickness 30 nm) are immersed in air ($n_0$). The Au and $Si_3N_4$ squared structures in Fig. 1 have dimensions 950 × 950 nm with periodic boundary conditions on the top, left, bottom and right boundaries (i.e., *x*-, *y*- boundaries). In so doing the structures are infinitely repeated along the *x*- and *y*- axes. Note that the large lateral size of the simulation domain (950 × 950 nm) has been chosen to minimize lattice resonances. An extremely fine mesh was then applied with maximum and minimum element sizes of $2.19 \times 10^{-7}$ m and $9.38 \times 10^{-9}$ m. Moreover, to further improve the mesh of the plasmonic structure, two extra automatic refinements were applied to the layer containing the plasmonic features (particles or apertures). Two ports (one at the front and one at the back of the structures) are implemented to apply the incident planewave and to record the transmitted signal, respectively. These ports are each placed at a distance 3000 nm, along the *z*-direction, away from the input (metal)/output (substrate) surfaces of the plasmonic structures. As explained before, the incident planewave is polarized along the *x*- or *y*-directions (transverse or parallel to the long axis of the plasmonic structures, respectively, see Fig. 1).

With this configuration, the incident planewave (indicated by the arrow labelled *Inc.* in Fig. 1) interacts with the plasmonic structures to excite LSP resonances. As it will be shown, illuminating the plasmonic particles and apertures with two different linear, orthogonal polarizations will excite different but complementary LSP resonant modes. To determine their spectral position, the reflected (labelled as *R*) and transmitted (labelled as *T*) radiation is measured. The magnitude of the transmitted and reflected signals as a function of frequency is calculated by recording the scattering parameters (*S*-parameters) at the input/output ports as defined by COMSOL Multiphysics®. The reflection and transmission spectra are then used for each of the planewave illuminated plasmonic structures to determine the spectral position of each LSP resonance.

**Localized surface plasmons (LSP) modes**

The schematic representations of the plasmonic particles and apertures under study are shown in the first column of Fig. 2 [Fig. 2a-d(i), top views at the surface of the plasmonic particles/apertures]. In this figure, the linear polarization of the incident signal used to illuminate the plasmonic structures is represented as a horizontal or a vertical arrow depending on the polarization direction of the incident electric field ($E_x$ or $E_y$, respectively). To begin with, the reflection and transmission for the plasmonic particles and apertures under vertical and horizontal polarization of the incident illumination are shown in the second column of Fig. 2 [Fig. 2a-d(ii)] as black (*R*) and red lines (*T*). As observed, there are spectral minima in the transmission for the plasmonic particles which almost match the spectral location of the minima of the reflected signals for the plasmonic apertures. These minima are indications of the excitation of LSP resonances in both structures. The existence of these LSP resonances appearing at almost the same location in the spectrum for the transmission coefficient of the plasmonic particles and the reflection of the plasmonic apertures under orthogonal polarization demonstrates that Babinet's principle is *approximately* but not completely valid in these plasmonic structures. This is an expected result given that the metallic layers are non-infinitely thin and the fact that the metals are not perfect electric conductors at optical frequencies, which are requirements for Babinet's principle to hold. As a quantitative example, consider the plasmonic particles illuminated by a planewave with a vertical $E_y$ polarization as shown in Fig. 2a(i). For this scenario, the LSP resonance [minimum transmission, red plot in Fig. 2a(ii)] is a longitudinal dipole bonding (LDB) LSP mode (obtained at frequency/wavelength/energy $f_0 \approx 259$ THz/$\lambda_0 \approx 1.16$ µm/$E_0 \approx 1.07$ eV). Now, for its complementary plasmonic version, i.e., apertures illuminated by a planewave with $E_x$ polarization [see Fig. 2b(i)] the LSP resonance appears at a frequency of ~250 THz ($\lambda_0 \approx 1.12$ µm/$E_0 \approx 1.03$ eV) and corresponds to a complementary longitudinal dipole bonding (cLDB) LSP mode [calculated from the minimum in reflection of Fig. 2b(ii), black plot]. As explained above, this is an expected result due to the fact that here realistic plasmonic structures are considered (instead of using ideal perfectly conducting metals with almost zero thickness) which produce the LSP resonances in the plasmonic particles and complementary plasmonic apertures at a similar but not exactly the same spectral positions. For completeness, let us also consider the scenarios shown in Fig. 2c,d(i) for an $E_x$ and $E_y$ polarized incident illumination of the plasmonic particles and apertures, respectively. In these cases, the LSP resonance obtained using the minimum of transmission and reflection for the plasmonic particles and apertures, respectively, are a transverse dipole antibonding (TDA) LSP mode ($f_0 \approx 290$ THz/$\lambda_0 \approx 1.03$ µm/$E_0 \approx 1.20$ eV) [Fig. 2c(ii)] and a complementary transverse dipole antibonding (cTDA) LSP mode ($f_0 \approx 280$ THz /$\lambda_0 \approx 1.07$ µm/$E_0 \approx 1.16$ eV) [Fig. 2d(ii)], respectively. Again, this demonstrates how the LSP resonances occur in similar but not the same spectral positions. For the mode labeling see Ref. [30].

Once the spectral location of the excited LSP modes for the plasmonic particles and apertures have been studied, we can now discuss the nature and similar properties of these plasmonic resonances in terms of their field enhancement distribution (from now on just referred to as field distributions). Here enhancement is defined as the ratio between the spatial distribution of the electric or magnetic field with and without using the plasmonic structures. As demonstrated in [40], [44], [74], the presence of two plasmonic features (cylinders in our case) will generate LSP modes which are the result of hybridization of individual LSP modes excited in each plasmonic particle or aperture. However, as we are dealing with complementary structures, while the nature of the hybridized modes for both plasmonic structures are different (as it will be shown below) their electric ($E$) and magnetic ($H$) field distributions will be complementary, following Babinet's principle.

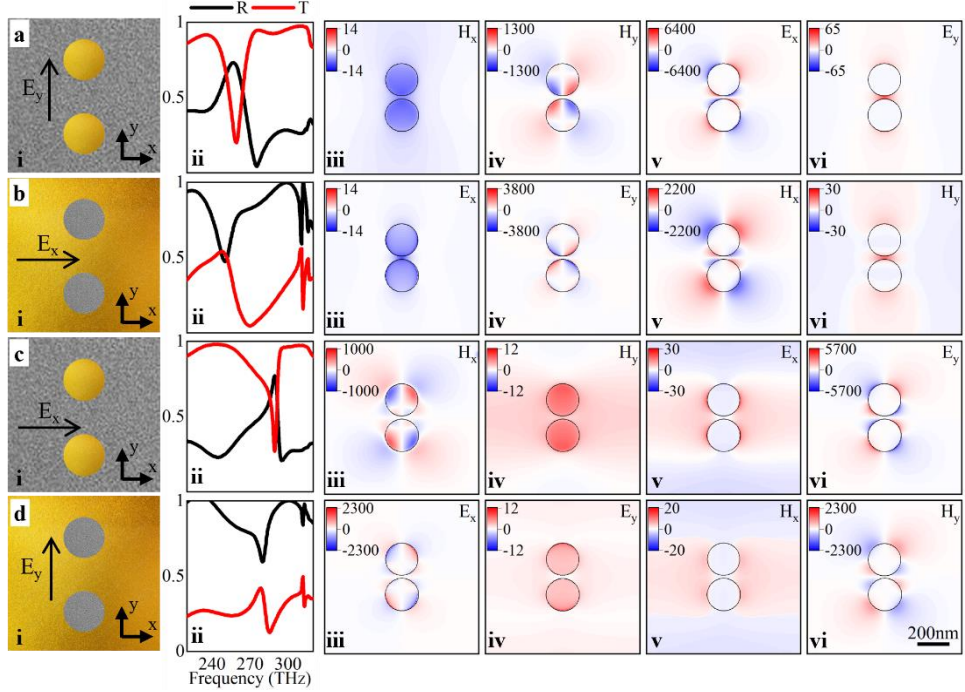

**Fig. 2| Field distribution of plasmonic particles and apertures.** Each row in this figure represents: (a,c) The results considering the plasmonic particles illuminated with a linearly $E_y$ or $E_x$ polarized planewave, respectively. (b,d) The results considering the plasmonic apertures illuminated with a linearly $E_x$ or $E_y$ polarized planewave, respectively. The panels along the rows (a,c), i.e., the plasmonic particles, are organized as follows: (i) Two-dimensional (2D) schematic representation on the *xy*-plane of the plasmonic particles showing the direction of the linearly polarized incident planewave illuminating them, (ii) reflection (black solid line) and transmission (red solid line) spectra, (iii) $H_x$, (iv) $H_y$, (v) $E_x$, and (vi) $E_y$ field enhancements of the LSP resonances: longitudinal dipole bonding (LDB) LSP mode ($f_0 \approx 259$ THz/$\lambda_0 \approx 1.16$ µm/$E_0 \approx 1.07$ eV) for (a) and a transverse dipole (TD) LSP mode ($f_0 \approx 290$ THz/$\lambda_0 \approx 1.03$µm/$E_0 \approx 1.20$ eV) for (c). The field enhancements are calculated as the ratio between the spatial distributions of the electric or magnetic field with and without using the plasmonic structures. Similarly, for panels (b,d) corresponding to the plasmonic apertures, the panels along the rows are organized as follows: (i) 2D schematic, (ii) reflection and transmission spectra , (iii) $E_x$, (iv) $E_y$, (v) $H_x$, and (vi) $H_y$ field enhancements of the LSP resonances: complementary longitudinal dipole bonding (cLDB) LSP mode ($f_0 \approx 250$ THz/$\lambda_0 \approx 1.12$ µm/$E_0 \approx 1.03$ eV) for (b) and a complementary transverse dipole (cTD) LSP mode ($f_0 \approx 280$ THz /$\lambda_0 \approx 1.07$ µm/$E_0 \approx 1.16$ eV)  for (d).

Let us discuss the complementary field distributions for both plasmonic structures (a discussion regarding the distribution of charge mapping the different hybridized LSP modes will be presented in the next sections). The $E$- and $H$-field distributions of the plasmonic LSP resonances, calculated at the surface of the plasmonic features, are shown in Fig. 2a-d(iii-vi). For the plasmonic particles (first and third row from Fig. 2) the field distributions are calculated for the LDB and TDA LSP modes, respectively, i.e., the frequency of minimum transmission. For the plasmonic apertures (second and forth rows from Fig. 2 corresponding to the cLDB and cTDA LSP modes, respectively) the LSP resonances are calculated at the minimum of reflection, as explained above. With this configuration, let us first compare the results shown in Fig. 2a,b. As observed, the horizontal ($H_x$) [Fig. 2a(iii)] and vertical ($H_y$) [Fig. 2a(iv)] components of the ***H***-field for the plasmonic particles under vertical polarization of the incident planewave resembles the horizontal ($E_x$) [Fig. 2b(iii)] and vertical ($E_y$) [Fig. 2b(iv)] components of the ***E***-field distribution of the plasmonic apertures, respectively. This is also evident when comparing the $E_x$ [Fig. 2a(v)] and $E_y$ [Fig. 2a(vi)] field distributions of the LSP resonance for the plasmonic particles with the $H_x$ [Fig. 2b(v)] and $H_y$ [Fig. 2b(vi)] field distributions of the plasmonic apertures. Finally, the same comparison can also be applied to the field distributions shown in Fig. 2c,d, demonstrating how the LSP modes are complementary, as it should happen when considering Babinet's principle [75]. Quantitatively, note that from Fig. 2, the components for the ***E*** and ***H*** fields have different magnitudes, in agreement with [67].

**Fabrication and experimental comparison**

Both the plasmonic particles and apertures were prepared using a standard focused ion beam (FIB) lithography process [76] which produces high-quality polycrystalline plasmonic antennas fully equivalent to monocrystalline ones [77]. First, a 30-nm-thick gold layer was deposited by magnetron sputtering on a standard silicon nitride membrane for transmission electron microscopy (TEM) with lateral dimensions of $250 \times 250$ μm$^2$ and a thickness of 30 nm. Second, the dimers were fabricated by FIB milling (using Ga+ ions at 30 keV) of the gold layer in a dual beam microscopy system FEI Helios. The particles were set in the center of a gold-free rectangular area with the size of $3\times2$ μm$^2$ to prevent any undesired interaction with the surrounding gold frame.

To characterize the morphology and optical properties (i.e., localized surface plasmon modes) of the dimers, the sample was analyzed by scanning transmission electron microscopy (STEM) in combination with electron energy loss spectroscopy (EELS) in a transmission electron microscope FEI Titan in monochromated scanning regime. The parameters were set as follows: the electron energy of 300 keV, with the beam current around 100 pA, the convergence semi-angle of 10 mrad, and the collection semi-angle of 56 mrad. The full width at half maximum of the zero-loss peak read 0.14 eV, and this value roughly represents the spectral accuracy of EELS. Note that prior to the STEM-EELS measurements, the sample was cleaned in argon-oxygen plasma for 20 seconds to prevent the sample from carbon contamination evolution during the measurement.

Annular dark-field (ADF) STEM images of the fabricated plasmonic particles and apertures are shown in Fig. 3a,g, respectively. The real dimensions of the structures determined from ADF-STEM images slightly differ from the targeted ones (the cylinder diameter of 200 nm and the gap between the dimers of 20 nm). For the plasmonic particles (Fig. 3a) we obtained the cylinder diameter of 172 nm (both particles) and the gap of 50 nm, while the plasmonic apertures, shown in Fig. 3g, have the diameters of 210 nm (top aperture) and 200 nm (bottom aperture) and a gap of 20 nm.

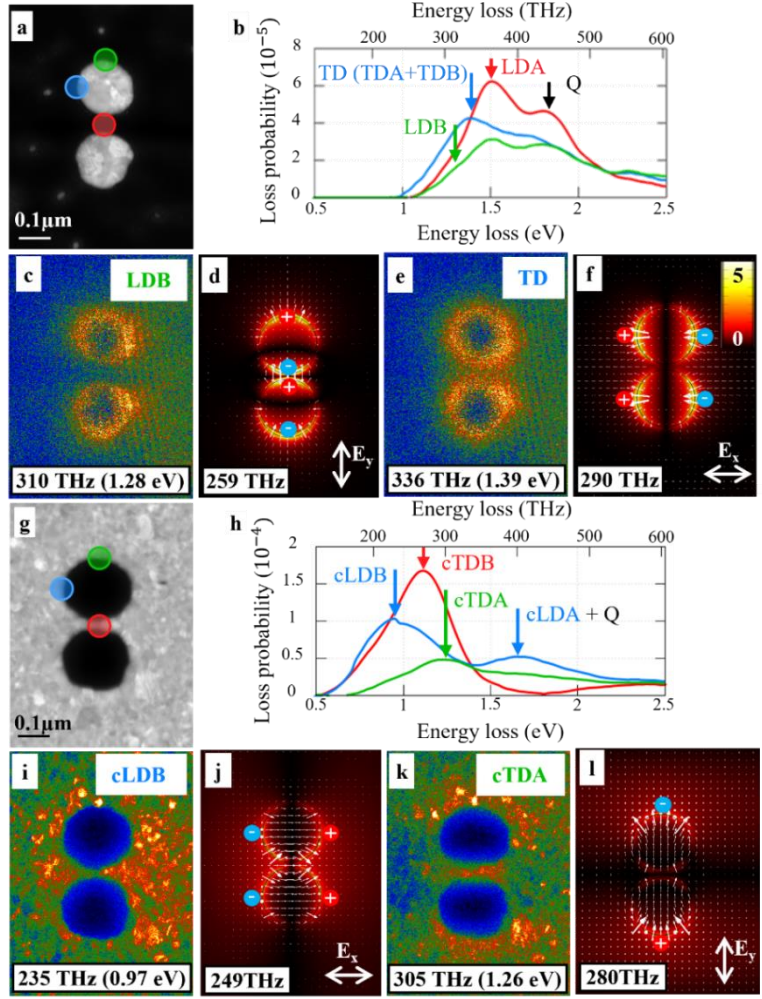

**Fig. 3 | STEM characterization of the plasmonic particle and aperture dimers**. (a,g) ADF-STEM images of a particle and aperture dimer, respectively. For the plasmonic particles, the real dimensions are: diameters $D_t$ = 172 nm (top particle) and $D_b$ = 172 nm (bottom particle) and the gap L = 50 nm, while for the plasmonic apertures, the real dimensions are $D_t$ = 210 nm (top aperture) and $D_b$ = 200 nm (bottom aperture) and L = 20 nm. (b,h) Loss probability (per the spectral range of 0.01 eV) for the plasmonic particles and apertures, respectively, measured at positions marked in panels (a,g). (c,e) Spatial maps of the loss probability for the plasmonic particles shown in (a) for a frequency of the LDB and TD modes, respectively. (d,f) Numerical results of the $|E_z|$ field at the surface of the dimers ($D_t$ = $D_b$ = D = 200 nm, L = 20 nm) illuminated by a planewave with $E_y$ polarisation at a frequency of the LDB mode and $E_x$ polarisation at a frequency of TDA mode, respectively, along with the corresponding electric field flux lines. In these panels, the "(+)" (red) and "(-)" (blue) symbols represent the surface charge distributions. (i,k) Same as (c,e) but for the plasmonic apertures at the frequency of the cLDB and cTDA modes, respectively. (j,l) Same as (d,f) when considering the complementary apertures using an $E_x$ polarised incident planewave at the frequency of cLDB mode and using a planewave with $E_y$ polarisation at a frequency of cTDA mode, respectively.

The LSP modes supported by plasmonic dimers are characterized by EELS. A beam of probing electrons is transmitted through the sample and an energy spectrum of inelastically scattered electrons is recorded. The spectrum is contributed also by the electrons that excited a LSP mode in a plasmonic dimer, and in turn decreased their energy by the energy of the LSP mode. The excitation of a LSP can be described within the framework of classical electrodynamics where the EM field induced by the LSP acts back on the electron [78], [79], resulting in the loss probability [67] :

$$\Gamma_{EELS}(\omega) = \frac{e}{\pi\hbar\omega} \int dt Re\{e^{-i\omega t}\boldsymbol{v} \cdot \boldsymbol{E}_{ind}[\boldsymbol{r}_e(t),\omega]\} \qquad (1)$$

where $e$ is the charge of an electron, ω the working angular frequency, $\hbar$ is the reduced Planck constant, $t$ is time and $\boldsymbol{E}_{ind}[\boldsymbol{r}_e(t),\omega]$ is the induced electric field from an electron moving with a velocity $\boldsymbol{v}$ at the electron position $\boldsymbol{r}_e(t)$. Now, if one considers an electron moving along the *z*-axis (perpendicular to the plasmonic particles or apertures) only the out-of-plane component of the electric field ($E_z$) of the plasmonic structures will interact with the probing electrons, making EELS insensitive to in-plane E-fields ($E_x$ and $E_y$). The LSP-related losses are proportional to the $E_z$ of the excited LSP resonances. The experimental EEL spectrum is further contributed by material-related bulk losses which are proportional to the thickness of the metallic structure. To isolate the LSP-related losses we performed the subtraction of the zero-loss peak and background and normalized the spectra.

Experimental EEL spectra are shown in Fig. 3b for the plasmonic particles and in Fig. 3h for the plasmonic apertures. The identification of the LSP modes in the plasmonic particle spectra follows the procedure described in Ref. [80] and for the interpretation of the plasmonic apertures spectra we utilize the electric-magnetic Babinet complementarity [30]. The spectral positions of all hybridized dipole modes [81] [i.e., LDB, transverse dipole bonding (TDB), transverse dipole anti-bonding (TDA), and longitudinal dipole antibonding (LDA)] supported by the plasmonic particles and Babinet-complementary modes supported by the plasmonic apertures (with abbreviations prefixed with c, i.e., cLDB, cTDB, cTDA, cLDA), are marked with arrows in Fig. 3b,h. We also utilize the label TD for unresolved TDB and TDA modes supported by the plasmonic particles and Q for the contribution of the quadrupole mode (disregarding its hybridization). In the following we focus on bright LSP modes with non-zero dipole electric moment as they couple to the EM planewave and are manifested in the transmission and reflection spectra of Fig. 2. They include LDB and TDA modes of the plasmonic particles and cLDB and cTDA modes of the plasmonic apertures. The other dipole LSP modes are dark [20], [28]. The spectral positions of the LSP modes are shown in Table 1. As observed, they correspond rather well to the spectral positions obtained from the calculated optical spectra and they also approximately follow Babinet's principle. The differences between the numerical and experimental location of the LSP resonances may be attributed to fabrication tolerances introducing differences in the dimensions between the fabricated and numerically simulated plasmonic structures. These may include, for instance, the fabricated plasmonic dimers being not completely cylindrical or the $Si_3N_4$ substrate not being perfectly 30 nm, among others. A study of the influence of these potential errors is shown in the supplementary materials. Moreover, as shown in Table 1, somewhat larger mode energies observed for the plasmonic particles can be attributed to the cylinder diameter being smaller than designed (by about 14 %). Based on the approximate linear dispersion relation which holds for gold plasmonic antennas (see e.g. [30] and Fig. 4 therein) we can qualitatively correct the spectral position by about 14 % (and for the aperture dimer by about 2.5 % upwards), obtaining even better agreement. Fig. 3c,e and Fig. 3i,k show the spatial maps of the experimental loss probability at the energy of the LDB and TDA modes (Fig. 3c,e, the plasmonic particles) and cLDB and cTDA modes (Fig. 3i,k, the plasmonic aperture). To reduce the noise, the loss probability maps are integrated over a spectral range of 0.1 eV centered around the mode energy.

*Table 1: Spectral positions of the bright dipole LSP modes supported by the plasmonic particles and the plasmonic apertures.*

|  | Experimental | Corrected | Calculated | | |
|---|---|---|---|---|---|
| Mode | $f_0$(THz) | $f_0$(THz) | $f_0$(THz) | $E_0$(eV) | $\lambda_0$($\mu m$) |
| **Particles** | | | | | |
| LDB | 310 | 267 | 259 | 1.07 | 1.16 |
| TDA | 336 | 289 | 290 | 1.20 | 1.03 |
| **Apertures** | | | | | |
| cLDB | 235 | 241 | 250 | 1.03 | 1.12 |
| cTDA | 305 | 312 | 280 | 1.16 | 1.07 |

To better understand the EELS results, the numerical simulation results using COMSOL Multiphysics® of the absolute value of the out-of-plane electric field ($|E_z|$) distribution along with the flux lines of the ***E***-field when the plasmonic structures are illuminated by a linearly polarized planewave (direction displayed by the white arrows) at the LSP resonant frequency [using the results from Fig. 2a-d(ii)] are shown in Fig. 3d,f for the LDB and TDA modes supported by the plasmonic particles and in Fig. 3j,l for the cLDB and cTDA modes supported by the plasmonic apertures. In these panels, the representation of accumulation of charges of each LSP mode is shown as red circles with "(+)" symbol and blue circles with "(-)" symbols, respectively. The position of the accumulated charge is determined by Gauss's law, as the strongest $|E_z|$ distribution will occur where charge has accumulated [30]. As observed, when the plasmonic particles are illuminated by an $E_y$ polarized planewave and an LDB mode is excited, as in Fig. 3d, there is a high $|E_z|$ field distribution and clear accumulation of charges at the top and bottom of each individual plasmonic dimer, confirming the assignment of the mode as LDB. On the other hand, when considering the $E_x$ polarization at a frequency of a TDA mode, (shown in Fig. 3f, the $|E_z|$ field distribution and therefore the charges are moved to the sides of the plasmonic dimers, a feature indeed corresponding to a TD LSP mode. For the plasmonic apertures illuminated by a $E_x$ polarized planewave, the numerical result of $|E_z|$ at the frequency of the cLDB mode is shown in Fig. 3j. In these results, bright spots can be clearly seen at the edges of the plasmonic apertures with a lower $|E_z|$ field distribution towards the gap between the plasmonic apertures. Meanwhile, when the plasmonic apertures are illuminated by a $E_y$ polarized planewave at the frequency of the cTDA mode (Fig. 3l), a high $|E_z|$ field distribution is observed at the top and bottom metallic regions of the apertures with a low $|E_z|$ field distribution at the gap suggesting that there is an accumulation of positive and negative charges making net zero charge at the gap, a feature of cTDA LSP modes. The experimental loss probability maps shall qualitatively agree with the calculated $|E_z|$ distribution, as predicted by Eq. 1. This is indeed confirmed for the plasmonic apertures. For the plasmonic particles, the large gap between the cylinders (50 nm) resulted in strong spectral overlap of all hybridized modes due to their smaller energy difference. This makes the visual comparison more challenging, while the qualitative agreement with numerical simulations is still present.

**Thin film sensing**

Let us now study the potential of both complementary plasmonic structures to sense variations of a nearby dielectric thin film. The schematic representations of the plasmonic particles and apertures explored in the previous sections with a thin dielectric film positioned atop (which will act as the analyte in our case) is shown in Fig. 4a,d, respectively. The refractive index, $n_a$, and thickness, $\delta_a$, of the analyte will be changed in order to shift the spectral position of the LSP resonances. This spectral shift is then mapped by recording the transmission and reflection spectra of the plasmonic structures (particles and apertures, respectively). This shift in the spectral position of the LSP resonances is shown in Fig. 4b,e for the plasmonic particles and apertures, respectively. Here, the plasmonic structures are illuminated with a planewave under $E_x$ (solid lines) and $E_y$ (dashed lines) polarization. The thickness of the analyte is then considered to be $\delta_a$ = 50 nm (black), 100 nm (red), 150 nm (blue) and 200 nm (orange) and its refractive index ($n_a$) is varied from 1.5 to 3 in steps of 0.5. The effect of the orthogonal polarizations of the incident planewave on the shift of the LSP resonances can be studied by looking at Fig. 4b,e. As it can be seen, the change of the resonant wavelength of the LSP mode for the plasmonic particles is more prominent when using $E_y$ polarization (dashed lines) while the spectral shift when using the plasmonic apertures is larger when illuminated by an orthogonally polarized planewave ($E_x$ polarization, solid lines). These are expected results due to the interaction between the analyte and the field distribution of the corresponding LSP modes for each plasmonic structure which produce high field concentrations near the plasmonic structures.

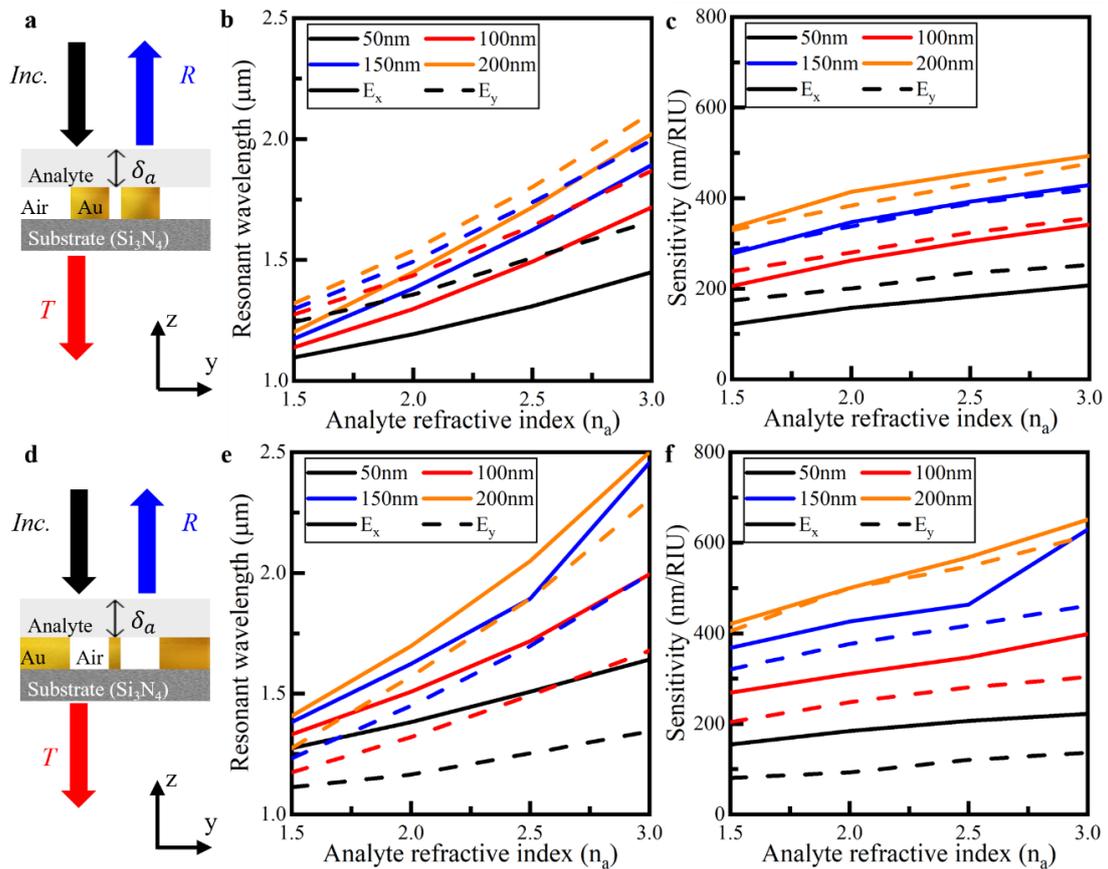

**Fig. 4| Resonant wavelength and sensitivity of the plasmonic structures with an analyte positioned atop.** 2D Schematic representation of the cross section on the *yz*-plane for (a) plasmonic particles and (d) apertures used to determine changes of a nearby thin film analyte. Resonant wavelengths of the LSP modes produced by the (b) plasmonic particles shown in (a), and (e) the plasmonic apertures from (d) when $n_a$ is changed from 1.5 to 3 in steps of 0.5 considering an incident planewave with $E_x$ (solid line) and $E_y$ (dashed line) polarisation. The thickness of the analyte is chosen to be: $\delta_a$ = 50 nm (black), $\delta_a$ = 100 nm (red), $\delta_a$ = 150 nm (blue) and $\delta_a$ = 200 nm (orange). Sensitivity of the (c) plasmonic particles and (f) apertures when illuminated by a planewave polarised in the $E_x$ (solid) and $E_y$ (dashed) direction for analyte thicknesses of $\delta_a$ = 50 nm (black), $\delta_a$ = 100 nm (red), $\delta_a$ = 150 nm (blue) and $\delta_a$ = 200 nm (orange).

To further evaluate the sensing features of the plasmonic structures, their sensitivity can also be calculated. The sensitivity is defined as the ratio between the change of the wavelength of the LSP resonances and the refractive index variation of the analyte, $S = \frac{\Delta\lambda}{\Delta n_a}, \left[\frac{nm}{RIU}\right]$, with RIU as refractive index unit [82]–[84]. The calculated results are shown in Fig. 4c,f for the plasmonic particles and apertures, respectively. As observed, the sensitivity of both plasmonic structures increases as $\delta_a$ and $n_a$ of the analyte increases, as expected, in line with the results discussed in Fig. 4b,e. Quantitatively, the sensitivity is increased from ~100 nm/RIU for both the plasmonic particles and apertures (when $\delta_a = 50$ nm and $n_a = 1.5$) up to ~450 nm/RIU and ~650 nm/RIU for the plasmonic particles and apertures, respectively (when $\delta_a = 200$ nm and $n_a = 3$). It is important to note that when the thickness of the analyte on top of the plasmonic particles is $\delta_a < 100$ nm there is a clear difference between values of sensitivities for the two orthogonal polarizations of the incident illumination, with sensitivities larger when using the $E_y$ polarization. For the plasmonic apertures, the sensitivity values are generally higher when illuminated by a planewave with $E_x$ polarization (as discussed before in terms of the wavelength shift). From these results with both plasmonic structures, their increased sensitivity depending on the polarization of the illumination is due to the excitation of a hotspot in the field distribution: ***E*-**field hotspot between the plasmonic particles and a complementary ***B***-field hotspot between the plasmonic apertures. Moreover, as the analyte touches the plasmonic apertures on all its surface, compared to the plasmonic particles where only the top surface of the particles is in contact to the analyte (see Fig. 4a,d), these field distribution hotspots will cause a more significant interaction of LSP modes with the analyte for the plasmonic apertures, increasing the shift in the resonant wavelength compared to the plasmonic particles (Fig. 4b,e) making them more sensitive to changes of the analyte as shown in Fig. 4c,f. The values of sensitivity shown here are in line to values found in the literature with similar structures [82], [85], [86]. Finally, it is important to note that when using thicker dielectrics ($\delta_a > 100$ nm) for both plasmonic particles and apertures, the sensitivity values obtained with both polarizations of the illuminating signal are similar (i.e. almost independent on the polarization). This is because of the interaction of the near field distribution of the LSP resonances with the dielectric analyte on top of the plasmonic structures: for thin dielectrics, the near field of the LSP mode interacts more with the whole analyte compared with the interaction when using thick dielectrics (evanescently decaying within the dielectric). Meaning that the change in sensitivity observed with thick dielectrics may be mainly influenced by the multiple reflections within the dielectric rather than the interaction with the LSP resonance of the plasmonic particles and apertures, producing similar values of sensitivities regardless of the polarization of the incident planewave.

**Complementary dielectric sensing: an alternative sensing approach**

In the previous section the performance of the proposed plasmonic particles and apertures was studied when working as thin dielectric sensors. For completeness, a final study was carried out to investigate the sensitivity of the plasmonic particles and apertures when, instead of using a thin film positioned atop the metallic structures, the dielectric analyte was used as part of the complementary plasmonic structures. A 2D schematic representation of the cross section of the cylindrical plasmonic particles is shown in Fig. 5a where it can be seen that the particles are immersed within the dielectric analyte (rather than air as in the previous sections) with both (analyte and particle) having the same thickness. Similarly, to study a complementary sensing approach, the plasmonic apertures are then filled with the dielectric analyte as it is schematically represented in Fig. 5f. With this configuration, the numerical results of the power enhancement on the *xy*- and *zy*-planes for the plasmonic particles in a dielectric analyte ($n_a = 1.5$) illuminated by an $E_y$ polarized planewave with a frequency of ~239 THz ($\lambda_0 \approx 1.25$ µm/$E_0 \approx 0.99$ eV) is shown in Fig. 5b,c, respectively. A power enhancement (defined as the ratio between the spatial power distribution of the with and without using the plasmonic structures) of ~20 is obtained at the gap of the plasmonic particle structure, showing a field hotspot. The power enhancement of the complementary plasmonic structure, double cylindrical dielectrics in a metallic sheet illuminated by an orthogonally polarized planewave at ~241 THz ($\lambda_0 \approx 1.24$ µm/$E_0 \approx 1.00$ eV) on the *xy*- and *zy*-plane is shown in Fig. 5g,h, respectively. Here, a field hotspot with a power enhancement of ~60 between the plasmonic apertures is observed as the LSPs are tightly bound to the Au bridge between the two apertures. As expected by Babinet's principle, both complementary structures have similar power enhancement distributions although the magnitude of the power enhancements differ, as discussed in the previous sections.

The effect of changing the refractive index $n_a$ of the dielectric analyte on the transmitted and reflected signals for the plasmonic particles and apertures, respectively, is shown in Fig. 5d,i. Here, the resonant frequency (minima of the spectra, dashed lines have been added to guide the eye) of the LSPs modes is red shifted as $n_a$ increases, as expected from Fig. 4b,e. The sensitivity values of the plasmonic particles and apertures are shown in Fig. 5e,j, for both orthogonal planewave polarizations: $E_x$ (red) and $E_y$ (black). The sensitivities of the plasmonic particles, Fig. 5a,e, as $n_a$ changes from 1.5 to 3 is found to vary from 196 to 253 nm/RIU under $E_y$ illumination and from 90 to 150 nm/RIU when illuminated by an $E_x$ polarized planewave. On the other hand, the sensitivity of the filled plasmonic apertures, shown in Fig. 5f,j ranges from 91 to 133 nm/RIU and 15 nm/RIU to 50 nm/RIU when using a planewave with $E_x$ and $E_y$ polarization, respectively. Similar to the thin dielectric analyte atop the structures studied in the previous section, having the analyte as part of the plasmonic particles and apertures also increases their sensitivity as the refractive index increases. However, by comparing the results from Fig. 4c,f and Fig 5e,j, it can be seen that using the analyte as a part of the plasmonic particles and apertures structures (Fig 5e,j) reduces the volume of analyte required to achieve sensitivity values of the same order of magnitude compared to those shown in Fig. 4c,f. From Fig. 5e,j, the plasmonic particles and apertures illuminated by an $E_y$ or $E_x$ polarized planewave, respectively, achieve sensitivities of up to ~250 nm/RIU and ~150 nm/RIU ($n_a = 3$), respectively, with analytes of thickness $\delta_a = 30$ nm. These results are of the same order of magnitude to the sensitivities achieved, ~200 nm/RIU, when using a thin dielectric of $\delta_a = 50$ nm on top of the plasmonic structures (Fig. 4c,f). By comparing the results from Fig. 4 and those from Fig. 5 one can also notice the following: for the plasmonic particles, it can be seen that the sensing performance is improved when the particles are immersed within the analyte (Fig. 5e) compared to the case when the $\delta_a = 50$ nm analyte is placed atop (Fig. 4c). The opposite occurs with the plasmonic apertures where the sensitivity is improved when the $\delta_a = 50$ nm analyte is placed atop the apertures (Fig. 4f) compared to when the analyte is filling the apertures (Fig. 5j). This demonstrates that both plasmonic structures have complementary performances which may be applied for dielectrics sensing with a configuration that could be chosen depending on the type of dielectric to be sensed. In the supplementary materials, we also discuss the sensitivity normalized to the volume of the analyte. These results may find applications in dielectric sensing and biosensing, among others.

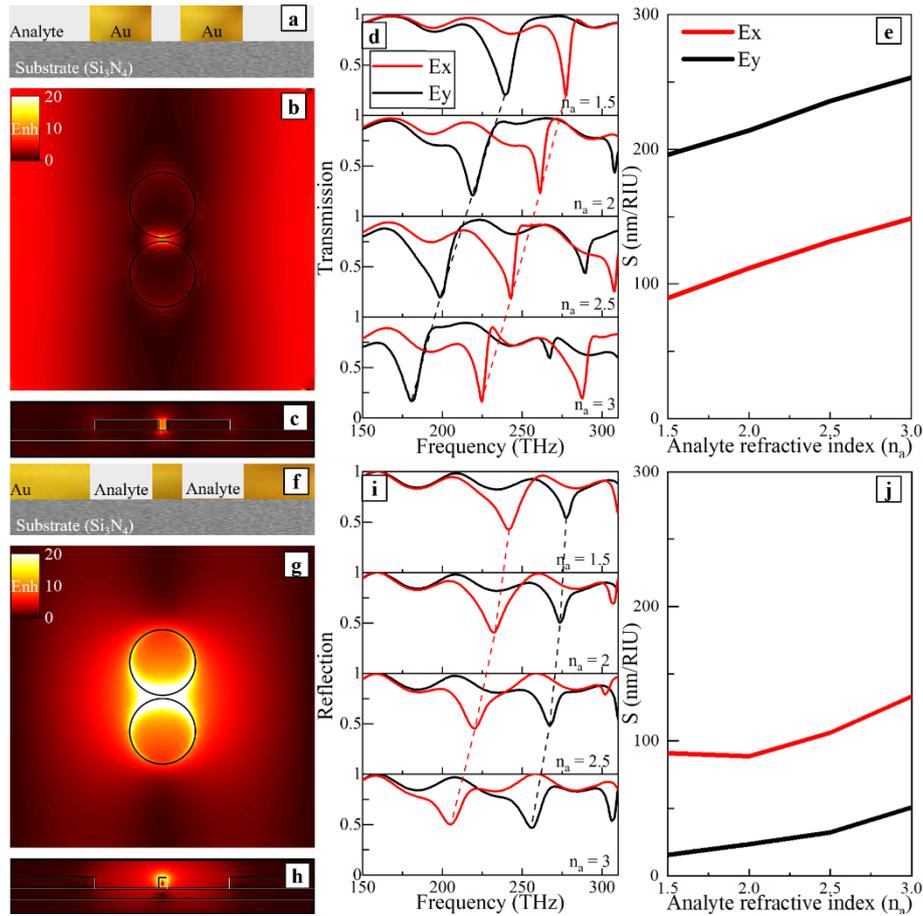

**Fig. 5| The plasmonic structures sensing an analyte as a part of the structures.** 2D cross section of (a) plasmonic particles and (f) apertures surrounded and filled by a dielectric analyte, respectively. Power enhancement on the (b,g) *xy*-plane and (c,h) *yz*-plane for the plasmonic particles and apertures shown in (a,f), respectively, considering an analyte ($n_a = 1.5$). The structures are illuminated with a planewave under $E_y$ or $E_x$ polarization, respectively. Note that the results in (g,h) have been saturated to use the same scale as (b,c) in order better observe the difference in power enhancement. (d) Transmission and (i) reflection spectra produced when illuminating the structures shown in (a,f), respectively, with an $E_x$ (red) and $E_y$ (black) polarized planewave when $n_a$ is changed from 1.5 to 3 in steps of 0.5 (panels from top to bottom). Dashed lines going through the minima of the spectra for each value of $n_a$ have been added to visualise the shift in the spectral location of the LSP resonance. Sensitivity of the (e) plasmonic particles and (j) apertures when they are illuminated by a planewave with orthogonal $E_x$ (red) and $E_y$ (black) polarizations.

## Conclusions

In this work, Babinet's principle of complementarity has been studied in the realm of plasmonics to develop sensors exploiting LSP resonances in complementary metal-dielectric structures. First, the Babinet principle in plasmonics was numerically analyzed by comparing the field distribution of the excited LSP resonances that exist in complementary cylindrical metallic particle-dimers and aperture-dimers in a metallic film. To further study them, these results were compared to experimentally fabricated plasmonic structures with the LSPs resonances mapped by EELS showing good agreement between them in terms of the excited LSP modes, their approximate complementary frequency where these LSP resonances occur and charge distribution. These structures were then used as dielectric sensors demonstrating how the LSP resonances can be shifted when using nearby dielectrics under two configurations: thin dielectric atop the plasmonic structures, or plasmonic particles/apertures immersed/filled with a thin dielectric. Complementary sensing performance was also demonstrated showing sensitivity values in the order of several ~100s nm/RIU for thin dielectrics of thicknesses as small as30 nm.

## Acknowledgements

V.P.-P. would like to thank the support of the Leverhulme Trust under the Leverhulme Trust Research Project Grant scheme (No. RPG-2020-316), and from Newcastle University (Newcastle University Research Fellowship). V.P-P. and J.A.R would like to thank the support from the Engineering and Physical Sciences Research Council (EPSRC) under the EPSRC DTP PhD scheme (EP/R51309X/1). V.K. and M.H. acknowledge the support from the Ministry of Education of the Czech Republic (project CzechNanoLab Research Infrastructure, No. LM2018110).

## Conflicts of interests

The authors declare no conflicts of interests.

## Author contributions

V.P.-P. and V.K. conceived and coordinated the research. J.R. and V.P.-P. performed all simulations. M.H. fabricated samples, characterized them with EELS and processed experimental data with assistance by V.K. All authors contributed to the discussion of the results and preparation of the manuscript.